# Room-Temperature Ferrimagnet with Frustrated Antiferroelectricity: Promising Candidate Toward Multiple State Memory

P. S. Wang and H. J. Xiang[*]

Key Laboratory of Computational Physical Sciences (Ministry of Education), State Key Laboratory of Surface Physics, and Department of Physics, Fudan University, Shanghai 200433, P. R. China

e-mail: hxiang@fudan.edu.cn

On the basis of first-principles calculations we show that the M-type hexaferrite $BaFe_{12}O_{19}$ exhibits frustrated antiferroelectricity associated with its trigonal bipyramidal $Fe^{3+}$ sites. The ferroelectric (FE) state of $BaFe_{12}O_{19}$, reachable by applying an external electric field to the antiferroelectric (AFE) state, can be made stable at room temperature by appropriate element substitution or strain engineering. Thus M-type hexaferrite, as a new type of multiferoic with coexistence of antiferroelectricity and ferrimagnetism, provide a basis for studying the phenomenon of frustrated antiferroelectricity and realizing multiple state memory devices.

The phenomenon of frustration, typically observed in the field of magnetism, is found in solids [1] and soft materials [2]. Prototype examples of geometrical spin frustration are provided by magnetic systems consisting of triangular or pyrochlore spin lattice with nearest-neighbor antiferromagnetic spin exchange. Frustrated magnetic systems can give rise to exotic phenomena such as quantum spin liquid [3] and spin-order induced ferroelectricity [4,5]. An important question in the field of ferroelectricity is whether geometrically frustrated antiferroelectricity exists or not. Currently, all well-known multiferroics (e.g., TbMnO$_3$, BiFeO$_3$) possess simultaneously ferroelectricity and antiferromagnetism. It is not clear whether antiferroelectricity can coexist with ferromagnetism.

Hexaferrite contains triangular lattices of $Fe^{3+}$ ions, which has been found to display intriguing magnetoelectric (ME) effects at room temperature and low magnetic fields (~ 0.01 T) [6,7,8,9]. The room-temperature insulating ferrimagnetic undoped M-type hexaferrite, $AFe_{12}O_{19}$ (A = Ca, Sr, Ba, Pb, etc.), was widely believed [10] to crystallize in the magnetoplumbite-type centrosymmetric structure (space group P6$_3$/mmc) with high-spin $Fe^{3+}$ ions in the octahedral (OCT) (12k, 4f and 2a), tetrahedral (TET) (4f) and trigonal bipyramidal (TBP) (2b) sites [Fig. 1(a)], see Fig. S1 of the supplementary material (SM) for details. The $Fe^{3+}$ ion position at the TBP site has been controversial. In the centrosymmetric structure, the TBP $Fe^{3+}$ ion lies on the equatorial plane (i.e., the local mirror plane) of the TBP FeO$_5$. The x-ray-diffraction study at room temperature [11] suggested that the $Fe^{3+}$ ion is displaced out of the equatorial mirror plane by about 0.16 Å, which was supported by a Mössbauer study [12]. Collomb *et al.* [13] carried out neutron-diffraction studies for a number of hexagonal ferrites at room temperature and 4.2 K. At room temperature, they found an even greater displacement (i.e., 0.26 Å) of the TBP site ions. However, their structure refinement at 4.2 K suggested a freezing of the $Fe^{3+}$ ion at the mirror-plane site, and this was supported by empirical rigid-ion model calculations [14]. In this work, we carry out a comprehensive first-principles study to resolve this controversy and reveal that the M-type hexaferrite $BaFe_{12}O_{19}$ exhibits frustrated antiferroelectricity, giving rise to both room-temperature polar order and strong ferrimagnetic order. Our work predicts that $BaFe_{12}O_{19}$ represents a first example of a new type of multiferroic possessing both antiferroelectricity and ferrimagnetism.

The presence of a structural instability in $BaFe_{12}O_{19}$ can be examined by computing its phonon dispersion within the density functional theory (see [15] for details). In $BaFe_{12}O_{19}$, the high-spin $Fe^{3+}$ ions (S = 5/2) order ferrimagnetically below 450 °C with 16 up-spin and eight down-spin $Fe^{3+}$ ions per unit cell, resulting in a net magnetization 20 $\mu_B$ per unit cell [10] (see Fig. S1). We first show that this ferrimagnetic state is the magnetic ground state by calculating the spin exchange parameters of $BaFe_{12}O_{19}$ (see SM), and then carry out phonon calculations for the ferrimagnetic spin ground state. Contrary to the conclusion of the empirical rigid-ion model [14], our calculations show the presence of two unstable modes in the whole Brillouin zone of the phonon dispersion [Fig. 1(b)], providing a clear evidence for the structural instability in $BaFe_{12}O_{19}$. Both unstable phonon modes at Γ are contributed mainly by the displacement of the $Fe^{3+}$ ions at the TBP sites along the c axis [Fig. 1(c)]. The lower (higher) frequency mode is associated with the in-phase (out-of-phase) vibration of the two TBP $Fe^{3+}$ ions in the unit cell. The eigenvectors of the two modes can be used to generate one FE and one AFE structure along the c axis. After performing structural relaxations, the FE and AFE structures become more stable than the centrosymmetric paraelectric (PE) structure by 4.3 meV/f.u. and 0.1 meV/f.u., respectively. This further evidences the structural instability of the TBP $Fe^{3+}$ ions at the local mirror-plane sites. In the FE structure, the TBP $Fe^{3+}$ ion moves out of the mirror plane by 0.19 Å, in good agreement with the experimental result (0.16 Å) [11].

We now investigate the interaction between the local dipoles caused by the displacements of the TBP $Fe^{3+}$ ions by considering the five different dipole arrangements (Fig. 2): (I) The 1×1 FE state in which all dipole moments aligned along the $c$ axis (i.e., $FE_{ab}$-$FE_c$ state); (II) The 1×1 $FE_{ab}$-$AFE_c$ state with the same $ab$-plane FE arrangement as the $FE_{ab}$-$FE_c$ state but with an antiparallel dipole moments between adjacent lattices along the $c$ axis; (III) The 2×1 $AFE_{ab}$-$FE_c$ state with the dipole moments aligned ferroelectrically along the $c$ axis and a chain-like AFE arrangement in the $ab$-plane; (IV) The $\sqrt{3}\times\sqrt{3}$ $FI_{ab}$-$FE_c$ state with the dipole moments aligned ferroelectrically along the $c$ axis and a two-up-one-down ferrielectric arrangement in the $ab$-plane; (V) The $\sqrt{3}\times\sqrt{3}$ $FI_{ab}$-$AFE_c$ state with the same $ab$-plane ferrielectric arrangement as the $FI_{ab}$-$FE_c$ state but with an antiparallel dipole moments between adjacent lattices along the $c$ axis. After structural relaxations, we obtain the relative energies of these five states summarized in Fig. 2, which shows that if two states have the same in-plane dipole arrangement, the state with a FE alignment of the dipoles along the $c$ axis has the lower energy [i.e., E($FE_{ab}$-$FE_c$) < E($FE_{ab}$-$AFE_c$), and E($FI_{ab}$-$FE_c$) < E($FI_{ab}$-$AFE_c$)]. On the other hand, the dipole moments prefer an AFE arrangement in the $ab$-plane [i.e., E($AFE_{ab}$-$FE_c$) < E($FI_{ab}$-$FE_c$) < E($FE_{ab}$-$FE_c$)]. This can be explained in terms of the dipole-dipole interaction (DDI),

$$E_{DDI}^{ij} = \frac{C}{r_{ij}^3}[\vec{p}_i \cdot \vec{p}_j - 3(\vec{p}_i \cdot \vec{e}_{ij})(\vec{p}_j \cdot \vec{e}_{ij})],$$

where $\vec{e}_{ij}$ is a unit vector parallel to the line joining the centers of the two dipoles, $r$ is the distance between two dipoles, $\vec{p}_i$ and $\vec{p}_j$, and $C$ is a constant related to the dielectric constant.

It can be easily seen that the two ferroelectrically aligned dipoles along the c axis has a lower DDI energy than that of an antiparallel dipole pair, while two dipoles with the dipole direction perpendicular to the distance vector tend to be antiparallel to each other (see the inset of Fig. 2). The energies ($E_{DFT}$) of the five states from the density functional theory (DFT) calculations are compared with their total DDI energies $E_{DDI} = \sum_{<i,j>} E_{DDI}^{ij}$ in Fig. 2, which reveals that the DFT results are very well described by the DDI model. This is due probably to the fact that the dipoles associated with the displacement of the TBP $Fe^{3+}$ ions are well separated from each other so that the short-range interactions between the dipoles are not important, unlike the case of traditional FE systems (e.g. $BaTiO_3$) [16].

We now examine the ground state and thermodynamic properties of $BaFe_{12}O_{19}$ using the DDI model since it closely reproduces the DFT results. Note that the dipoles form a hexagonal lattice and are always perpendicular to the *ab*-plane [i.e., along the *c* or –*c* direction (Ising-like)]. To find out the ground state configuration of the dipole arrangement, we adopt two approaches. One is to enumerate all the symmetrically nonequivalent configurations with the total number of dipoles no more than 12 in each supercell [17]. We find that the 2×1 AFE$_{ab}$-FE$_c$ state has the lowest DDI energy. The other approach is to perform parallel tempering Monte Carlo (MC) [18] simulations, which confirm that the 2×1 AFE$_{ab}$-FE$_c$ state (space group: Pnma) is the ground state and is consistent with the DFT result that it has the lowest energy among all five considered states. The ground state of the NN antiferromagnetic (AFM) Ising model on a triangular lattice is known to have a macroscopic degeneracy. The 6-fold degenerate 2×1 AFE$_{ab}$-FE$_c$ state has the lowest energy due to the long range nature of the DDI. As a matter of fact, a similar 2×1 chain-like AFM state is the ground state of the Ising model on a triangular lattice with AFM NN and AFM next NN interactions [19]. Although the AFE$_{ab}$-FE$_c$ state has the lowest energy, there are many low-lying excited states, and this affects the thermodynamic properties of BaFe$_{12}$O$_{19}$.

We perform parallel tempering MC simulations on a $10\times10\times4$ lattice using the DDI model to determine the thermodynamic properties of BaFe$_{12}$O$_{19}$. The effect of vibrational free energy is discussed in SM. Our calculations reveal that there is a sharp peak at around 3 K in the specific heat curve, indicating a long-range order of the dipoles (see Fig. 3). To characterize the phase transition, we compute the local correlation $\langle \vec{p}_i \cdot \vec{p}_j \rangle_{ab}$ between the NN dipole pair in the ab-plane and that $\langle \vec{p}_i \cdot \vec{p}_j \rangle_c$ between the NN dipole pair along the c axis. The $\langle \vec{p}_i \cdot \vec{p}_j \rangle_c$ becomes nonzero when the temperature is lowered below 15 K, and gradually increases to the maximum value of 1.0 when the transition temperature (3.0 K) is approached. The in-plane local correlation $\langle \vec{p}_i \cdot \vec{p}_j \rangle_{ab}$ becomes nonzero at much high temperature (about 300 K, not shown here). This is so because the distance (5.8 Å) between the NN dipoles in the plane is much shorter than that (11.5 Å) between the NN dipoles along the c axis, thus the in-plane DDI is much stronger. The in-plane correlation $\langle \vec{p}_i \cdot \vec{p}_j \rangle_{ab}$ saturates to $-\frac{1}{3}$ at the transition temperature, which is the smallest value that can be achieved in a 2D Ising triangular system. However, this does not mean that the system fully orders in the ground state below the transition temperature because there are many low-lying excited states with the same $\langle \vec{p}_i \cdot \vec{p}_j \rangle_{ab}$ and $\langle \vec{p}_i \cdot \vec{p}_j \rangle_c$ as does the ground state. For example, the $\sqrt{3}\times\sqrt{3}$ FI$_{ab}$-FE$_c$ is less stable than the ground state only by 0.15 meV per formula unit (FU). The existence of significant local correlations above the phase transition temperature is a typical signature of a frustrated system. The $2\times1$ AFE$_{ab}$-FE$_c$ ground state can be described as a modulated structure with a dipole modulation vector $\vec{q}$. There are three symmetrically equivalent modulation vectors: $\vec{q}_1 = (0.5, 0, 0)$, $\vec{q}_2 = (0, 0.5, 0)$, and $\vec{q}_3 = (-0.5, 0.5, 0)$. The order parameter can be chosen as the dipole structure factor $O = \frac{1}{N^2} \sum_{\vec{q}=\vec{q}_1, \vec{q}_2, \vec{q}_3} \sum_{ij} \vec{p}_i \cdot \vec{p}_j e^{i\vec{q} \cdot (\vec{r}_i - \vec{r}_j)}$, where $\vec{r}_i$ is the position of the *i*-th dipole, and N is the number of dipoles in the supercell. For the AFE$_{ab}$-FE$_c$ ground state, the order parameter is 1. Fig. 3 shows that this parameter starts to become non-zero only when the temperature is below the transition point, suggesting that the low temperature phase is the AFE$_{ab}$-FE$_c$ state.

We now compare our theoretical results with previous experiments. We find that the TBP $Fe^{3+}$ ion is displaced out of the equatorial mirror plane, which agrees with the x-ray-diffraction study [11] and the Mössbauer study [12]. The neutron-diffraction study by Collomb *et al.* [13] found a large displacement of the TBP $Fe^{3+}$ ion at room temperature, but suggested that the TBP $Fe^{3+}$ ion freezes at the mirror-plane site at 4.2 K. This is contrary to the usual phenomenon that symmetry lowers when temperature decreases as predicted by Landau's theory. The puzzling neutron-diffraction results may be due to the frustrated nature of the TBP $Fe^{3+}$ related dipoles and the fact that the TBP $Fe^{3+}$ ion is at the mirror-plane site on average in the AFE state. A future neutron-diffraction study at very low temperature (e.g., 1 K) may confirm the AFE ground state predicted in this work.

The above discussion shows the ground state of BaFe$_{12}$O$_{19}$ to be an AFE ferrimagnetic state. Our phonon calculations reveal that the FE ferrimagnetic state is metastable (see SM) because flipping a dipole needs to go through the high energy PE-like state. The metastability of the FE ferrimagnetic state suggests that BaFe$_{12}$O$_{19}$ is a promising candidate for realizing multiple state memory devices. We find that the FE BaFe$_{12}$O$_{19}$ has the same ferrimagnetic ground state and similar magnetic Curie temperature as PE BaFe$_{12}$O$_{19}$ (see SM), indicating that the spin-phonon coupling in this system is not very important. Our calculations show that the electric polarization (3.23 μC/cm$^2$) and magnetization (10 μ$_B$/FU) of the FE ferrimagnetic state in BaFe$_{12}$O$_{19}$ are both large. This is different from the usual single-phase multiferroics, for which either the electric polarization or the magnetization is small: For example, the polarization in Ba(Fe,Sc,Mg)$_{12}$O$_{19}$ is almost three-order of magnitude smaller [9].

However, the FE state of freestanding BaFe$_{12}$O$_{19}$ can be locally stable only at low temperature because the energy barrier for the dipole flip is about 2.1 meV/dipole. Therefore, we consider two ways of making this FE state stable at room temperature. One way is to replace some of the TBP Fe$^{3+}$ ions by other +3 ions. We examine the stability of the FE state by replacing the TBP Fe$^{3+}$ ions with nonmagnetic ions such as Al$^{3+}$, Ga$^{3+}$, Sc$^{3+}$, and In$^{3+}$. Fig. 4(a) shows that the FE state becomes more stable by replacing the TBP Fe$^{3+}$ ions with Al$^{3+}$ or Ga$^{3+}$ ions, but becomes less stable when the TBP Fe$^{3+}$ ions are replaced with Sc$^{3+}$ and In$^{3+}$ ions. This is explained by considering the ionic radii of the +3 ions. The interaction between an inert M$^{3+}$ cation and an O$^{2-}$ anion consists of the attractive Coulomb electrostatic interaction and the short-range Pauli repulsion between core electrons [20]. For a linear O-M-O arrangement, the Coulomb interaction favors a FE-like asymmetric arrangement with two different M-O bond lengths, while the Pauli repulsion favors a PE-like centrosymmetric arrangement. If the size of the metal ion is small, the Coulomb interaction will stabilize the FE-like arrangement. Otherwise, the PE-like arrangement becomes more stable. The ionic radii [21] increase in the order Al$^{3+}$ < Ga$^{3+}$ < Fe$^{3+}$ < Sc$^{3+}$ < In$^{3+}$, suggesting that the stability of the FE state follows the relationship Al$^{3+}$>Ga$^{3+}$>Fe$^{3+}$>Sc$^{3+}$>In$^{3+}$, in good agreement with our first-principles results [see Fig. 4(a)]. Thus, the FE state of BaFe$_{12}$O$_{19}$ can be made more stable at a higher temperature if the TBP Fe$^{3+}$ ions can be selectively replaced with smaller cations Al$^{3+}$ and Ga$^{3+}$.

The other way of stabilizing the FE state at room temperature is to apply compressive epitaxial strain [Fig. 4(b)]. An in-plane compressive epitaxial strain makes the TBP FeO$_5$ elongated along the c axis, so the distance between the Fe ion and the apical O ion becomes longer than the sum of the ionic radius, which enhances the FE distortion of the TBP Fe$^{3+}$ site. The stability of the FE state increases with decreasing the in-plane lattice constant. For example, if BaFe$_{12}$O$_{19}$ is grown on CaFe$_{12}$O$_{19}$ (which introduces a 2% compressive strain), the FE state is more stable than the PE state by about 26 meV/dipole, close to the room temperature energy scale. Our molecular dynamics [22] simulation shows that the FE state at a 5% compressive strain is stable at least up to room temperature: The TBP Fe$^{3+}$ ions stay at the original positions of an initial FE state after a 5 ps simulation at 300 K. As can also be seen from Fig. S7 of SM, the FE state becomes even more stable than the AFE state when the compressive strain is larger than 4% possibility due to the strain-polarization coupling. We also calculate the energy barrier from the FE state to the AFE state by using the climbing image nudged elastic band method [23]. As shown in Fig. 5, the barrier is 1.26 meV/f.u. in the case of zero strain, while the barrier is increased to 116.14 meV/f.u. at 5% compressive strain. Thus, the FE state could become stable both thermodynamically and kinetically at large compressive strain.

The above discussion suggests the possibility of realizing room temperature four-state memory devices [24] by selectively doping BaFe$_{12}$O$_{19}$ with smaller +3 ions at the TBP Fe$^{3+}$ sites or by growing BaFe$_{12}$O$_{19}$ on the hexagonal substrate with a small lattice constant. Gajek *et al.* successfully demonstrated [25] that a multiferroic tunnelling junction made by the FE and ferromagnetic La$_{0.1}$Bi$_{0.9}$MnO$_3$ can act as a four-state resistive memory system although its magnetic Curie temperature (105 K) is well below the room temperature. Our work may provide new recipes toward realizing a room-temperature four-state memory device.

The precise definition of antiferroelectricity is in general more subtle than for antiferromagnets and has not reach consensus. In this work, antiferroelectricity refers to the case where the ground state contains anti-parallel aligned dipole moments and there is a ferroelectric (FE) low-lying excited state. Recently, Rabe proposed the following definition [26]: *"an antiferroelectric is like a ferroelectric in that its structure is obtained through distortion of a nonpolar high-symmetry reference phase; for ferroelectrics, the distortion is polar, while for antiferroelectrics it is nonpolar. However, not all nonpolar phases thus obtained are antiferroelectric: in addition, there must be an alternative low-energy ferroelectric phase obtained by a polar distortion of the same high-symmetry reference structure, and an applied electric field must induce a first-order transition from the antiferroelectric phase to this ferroelectric phase, producing a characteristic P-E double-hysteresis loop."* For the system $BaFe_{12}O_{19}$ studied in this work, it satisfies almost all the above conditions. Currently, it is not clear whether a double hysteresis loop exists in $BaFe_{12}O_{19}$. The exact shape of the P-E curve depends on the temperature and the barrier between the FE state and the AFE state. If the FE state is metastable and the barrier is higher than the thermal energy ($k_BT$) at a given temperature T, then the zero-field polarization after a pooling process may be non-zero. Otherwise, there may be a P-E double-hysteresis loop. Some other factors such as domain pinning, and/or defects may also change the shape of the loop.

In summary, the M-type hexaferrite $BaFe_{12}O_{19}$ exhibits frustrated antiferroelectricity due to the local dipole moments arising from its TBP $Fe^{3+}$ ion sites, and is a novel multiferroic with both ferrimagnetic order and antiferroelectric order. The M-type hexaferrites are expected to provide a basis for realizing room-temperature multiple state memory devices.

We thank Professor M.-H. Whangbo for invaluable discussions. Work was supported by

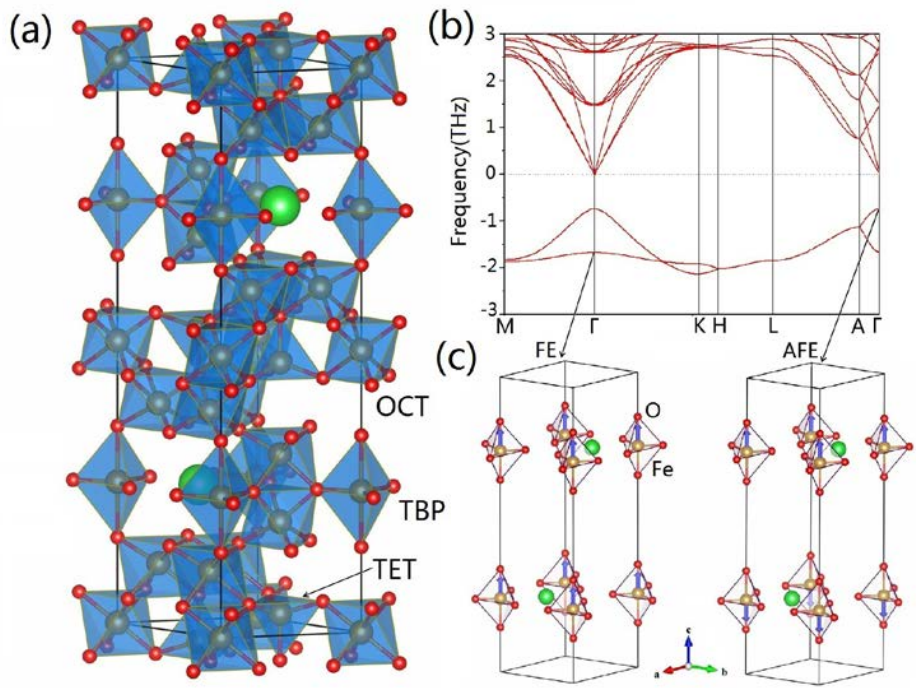

FIG. 1 (a) A polyhedral representation of the crystal structure of $BaFe_{12}O_{19}$. (b) The phonon dispersion relations calculated for $BaFe_{12}O_{19}$ by PBE+U calculations. For clarity, we only show the phonon branches below 3 THz. (c) The ionic displacements of the FE and AFE phonon modes at Γ calculated for $BaFe_{12}O_{19}$ by PBE+U calculations. For clarity, only the TBP $Fe^{3+}$ ions, their neighboring $O^{2-}$ ions, and the $Ba^{2+}$ ions are displayed.

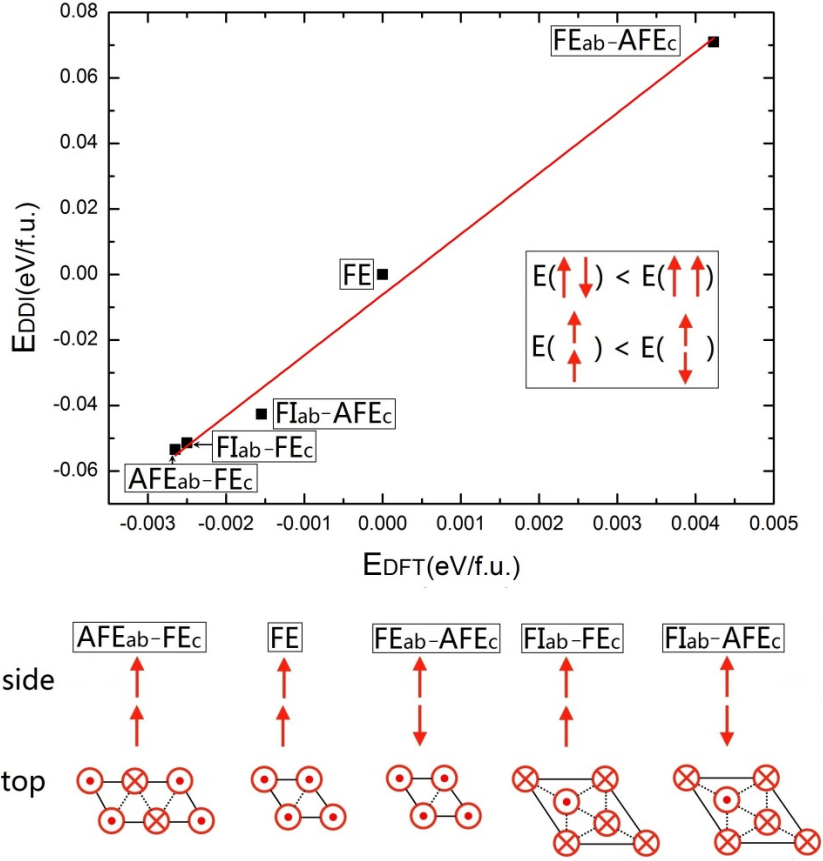

FIG. 2 **Comparison of the results from the DDI model with those from the DFT calculations.** The top panel plots the energies from the DDI model against the DFT energies for five different configurations (i.e., FE, AFE$_{ab}$-FE$_c$, FE$_{ab}$-AFE$_c$, FI$_{ab}$-FE$_c$, FI$_{ab}$-AFE$_c$) of the local dipoles. The zero energy reference is taken to be that of the FE state. A straight line from the linear fitting ($E_{DDI} = 18.5 E_{DFT} + c$) is also shown. The inset in the top panel schematically illustrates the idea that the two in-plane dipoles tend to be antiparallel to each other, while two dipoles along the c axis tend to be parallel to each other. The five dipole configurations are shown in the bottom panel. For the side view, we only indicate that whether the dipole

arrangement along c is AFE or FE. For the top view, ⊗ (⊙) represents the dipole along c (-c). The NN in-plane dipoles are connected by dashed lines. The solid lines denote the in-plane unit cell of the dipole configurations.

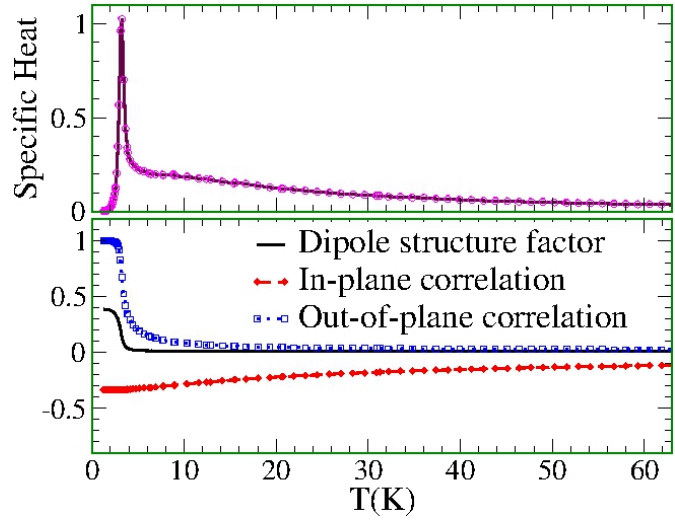

FIG. 3 **Thermodynamic properties of the dipoles in BaFe$_{12}$O$_{19}$ from the parallel tempering Monte Carlo simulations.** The specific heat curve shown in the top panel indicates a long range order at around 3.0 K. The bottom panel shows the spin structure factor (i.e, the order parameter O defined in the main text) and local dipole correlations (i.e., in-plane correlation $\langle \vec{p}_i \cdot \vec{p}_j \rangle_{ab}$ and out-of-plane correlation $\langle \vec{p}_i \cdot \vec{p}_j \rangle_c$) as a function of temperature.

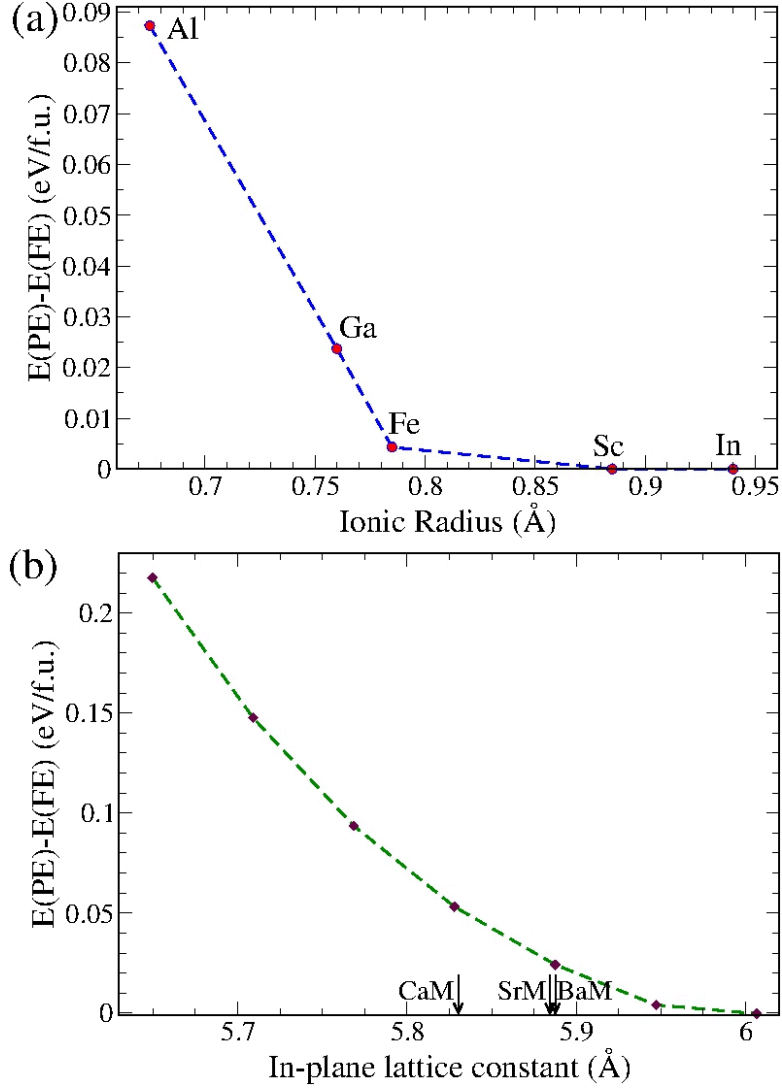

FIG. 4 **Stability of the FE distortion.** (a) The energy difference between the PE and FE state as a function of the radius of the +3 ion at the TBP $Fe^{3+}$ ion sites. (b) The energy difference between the PE and FE state as a function of the in-plane lattice constant. The out-of-plane lattice vector is fully optimized. The experimental in-plane lattice constants [10,36] of $AFe_{12}O_{19}$ with A = Ca, Sr, Ba are denoted by arrows.

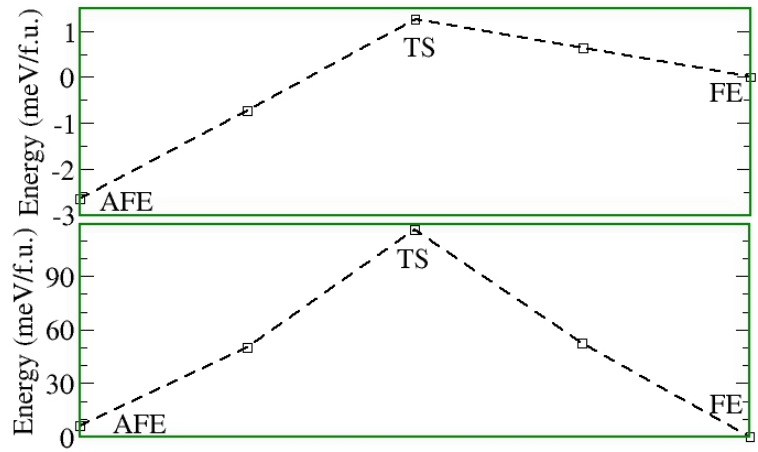

FIG. 5 **Energy barrier from the FE state to AFE (i.e., AFE$_{ab}$-FE$_c$) state.** Upper panel shows the case for the zero-strain case, while lower panel shows the case when the compressive epitaxial strain is 5%.